# Reproduction of Interstellar Infrared Spectrum of Reflection Nebula NGC2023 By A Hydrocarbon Pentagon-Hexagon Combined Molecule

NORIO OTA

Graduate School of Pure and Applied Sciences, University of Tsukuba,
1-1-1 Tenoudai Tsukuba-city 305-8571, Japan;   n-otajitaku@nifty.com

Reflection Nebula NGC2023 shows specific interstellar infrared spectrum due to polycyclic aromatic hydrocarbon (PAH) in a wide wavelength range from 5 to 20 micrometer. By our previous quantum chemistry calculation, it was suggested that a molecule group having hydrocarbon pentagon-hexagon combined skeleton could reproduce ubiquitous interstellar infrared spectrum. In this paper, observed NGC2023 spectrum was compared in detail with such carrier candidates. First model molecule was di-cation $(C_{23}H_{12})^{2+}$ with two hydrocarbon pentagons combined with five hexagons. Observed strong infrared bands were 6.2, 7.7, 8.6, and 11.2 micrometer. Whereas, calculated strong peaks were 6.4, 7.5, 7.7, 8.5, and 11.2 micrometer. Observed weaker bands from 10 to 15 micrometer were 11.0, 12.0, 12.7, 13.5, and 14.2 micrometer, which were reproduced well by computed bands as 10.9, 12.0, 12.6, 13.6, and 14.0 micrometer. From 15 to 20 micrometer, observed 15.8, 16.4, 17.4, 17.8, and 19.0 micrometer were correlated with calculated 15.6, 16.5, 17.2, 18.2, and 18.8 micrometer. It should be noted that we could successfully reproduce interstellar infrared spectrum by applying a single molecule. Second model molecule was $(C_{12}H_8)^{3+}$ with one pentagon combined with two hexagons. Again, observed strong bands at 6.2, 7.7, 8.6, 11.2 and 12.7 micrometer were successfully computed as 6.3, 7.4, 7.7, 8.6, 11.1, and 12.8  micrometer. It was concluded that by introducing hydrocarbon pentagon-hexagon combined ionized molecules, interstellar PAH oriented infrared spectrum could be successfully reproduced.

## 1, INTRODUCTION

Interstellar infrared spectrum (IR) due to polycyclic aromatic hydrocarbon (PAH) was ubiquitously observed in many astronomical dust clouds from 3 to 20μm (Boersma et al. 2013, 2014). However, any single PAH molecule or related species showing universal infrared spectrum has not yet been identified to date. Identification is essentially important to search chemical evolution step of organics and to study material building block of creation of life in the universe (Ota 2016). In 2014, accompanying a material study of void induced graphene sheet (Ota 2014a), it was incidentally founded that void induced graphene molecule $(C_{23}H_{12})^{2+}$ shows very similar infrared spectrum with interstellar observed one (Ota 2014b, 2015a). This molecule contains two hydrocarbon pentagons combined with five hexagons. Among observed eleven infrared bands, quantum chemical calculation suggested best agreement with eight major bands. This may lead to an identification of specific carrier molecule for interstellar infrared emission. After that,many PAH molecules were test for identification. Simple molecule $(C_{12}H_8)^{3+}$ had also show good coincidence with observed bands, which configuration was one hydrocarbon pentagon combined with two hexagons (Ota, 2015b).

Next question was how such hydrocarbon pentagon-hexagon molecules were created in space. I traced a history of star (Ota 2017), that is, nucleation of graphene after supernova (Nozawa et al. 2003, 2006), void creation by high speed proton, hydrogenation by low speed proton, and ionization by high energy photon. Energy diagram of each step was calculated resulting central star mass estimation to be 4~7 times heavier than our sun.

Very recently, Els Peeters et al. opened observed interstellar infrared spectrum of reflection nebula NGC2023 (Peeters 2017), which is one of the largest reflection nebula (4 light years diameter) showing detailed interstellar infrared spectrum due to PAH in a wide wavelength range from 5 to 20 micrometer. It's a very nice astronomical target to check general availability of modeled carrier. This paper compares detailed infrared spectrum between NGC2023 and model molecules. We could find essential coincidence with each other.

## 2, MODEL MOLECULES AND CALCULATION METHOD

In this study, two featured molecules $(C_{23}H_{12})^{2+}$ and $(C_{12}H_8)^{3+}$ are modeled as interstellar infrared spectrum carrier molecules. Supposed chemical evolution step is illustrated in Figure 1. After supernova (a), during expansion of helium sphere of onion like star internal structure, there happens super-cooling and nucleation of nano-carbon. Typical nano-carbon is graphene molecule (Ota 2017). Model molecule ($C_{24}$) has seven carbon hexagons as shown in column (A). By collision with interstellar gas cloud, nucleated graphene was sputtered by high speed proton $H^+$, which brings a molecular void inside of graphene as ($C_{23}$) in (b). Molecular configuration was immediately transformed to a quantum mechanically stable one having two carbon pentagons combined with five hexagons (c). Next step is hydrogenation by low speed proton modified to ($C_{23}H_{12}$) as like in (d). Final step is photoionization (e) by a central star illumination (Ota 2017). Interstellar molecules become cation as like $(C_{23}H_{12})^{n+}$ (n=1,2,3…). Similar modeling was tried on a smaller molecule ($C_{13}$) as shown in column (B), finally modified to $(C_{12}H_8)^{n+}$.

Calculation method is as follows. We have to obtain total energy, optimized atom configuration, and infrared vibrational mode frequency and strength depend on a given initial atomic configuration, charge and spin state Sz. Density functional theory (DFT) with unrestricted B3LYP functional was applied utilizing Gaussian09 package (Frisch et al. 2009, 1984) employing an atomic orbital 6-31G basis set. The first step calculation is to obtain the self-consistent energy, optimized atomic configuration and spin density. Required convergence on the root mean square density matrix was less than $10^{-8}$ within 128 cycles. Based on such optimized results, harmonic vibrational frequency and strength was calculated. Vibration strength is obtained as molar absorption coefficient ε (km/mol.). Comparing DFT harmonic wavenumber $N_{DFT}$ (cm$^{-1}$) with experimental data, a single scale factor 0.965 was used (Ota 2015b). Concerning a redshift for the anharmonic correction, in this paper we did not apply any correction to avoid over estimation in a wide wavelength from 2 to 30 micrometer.

Corrected wave number N is obtained simply by N (cm$^{-1}$) = $N_{DFT}$ (cm$^{-1}$) x 0.965.

Wavelength λ is obtained by λ (micrometer) = 10000/N(cm$^{-1}$).

Reproduced IR spectrum was illustrated in a figure by a decomposed Gaussian profile with FWHM=4cm$^{-1}$.

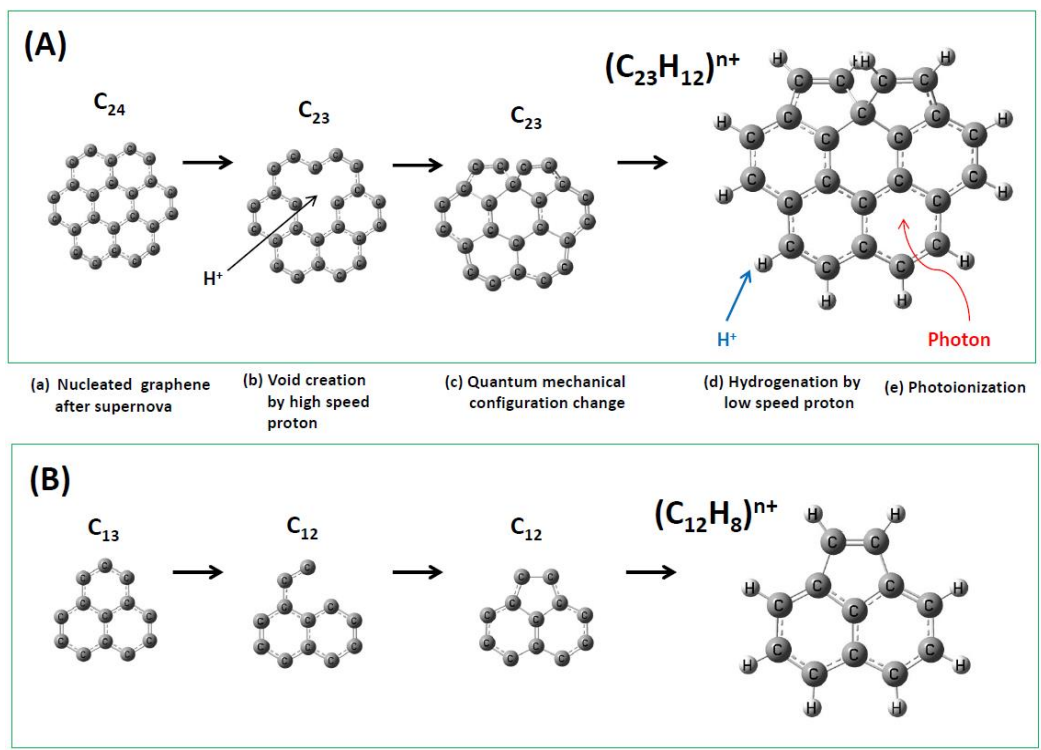

Figure 1, Chemical evolution model on typical carbon molecules. Nucleated graphene after supernova (a) would be sputtered by high speed proton (b) to make a void, which configuration is immediately changed to carbon pentagon-hexagon combined one (c). By slow speed proton sputtering, there occurs hydrogenation (d). Final step is photoionization by high energy photon irradiation (e).

## 3, CALCULATED INFRARED SPECTRUM OF $(C_{23}H_{12})^{2+}$

Calculated infrared spectrum of $(C_{23}H_{12})^{2+}$ was illustrated in a bottom of Figure 2, compared with observed NGC2023 (Peeters 2017) in top. Dotted lines are observed strong IR and arrow marks weaker one. Also, each wavelength were numerically compared in Table 1.

(1) Strong bands

  **Observed 6.2 $\mu$m band**: calculated value was 6.4 $\mu$m, red shifted by 0.2 $\mu$m. Vibrational mode is C-C stretching.

  **Observed 7.7 $\mu$m band:** Calculated band was split into 7.5 and 7.7 $\mu$m, which may be observed by single strong band. Mode is C-C stretching combined with C-H in-plane bending.

  **Observed 8.6 $\mu$m band:** Calculated value was 8.5 $\mu$m. Mode is C-C stretching and C-H in-plane bending.

  **Observed 11.2 $\mu$m band:** Calculated value was the same value. Mode is C-H out-of-plane bending, C-C stretching.

  Calculated 3.2 $\mu$m band: Well observed ubiquitous band. Mode is C-H stretching at pentagons.

(2) Accidental overlap with atomic ion emission

  Calculated 7.1 $\mu$m band: Accidental overlap with sharp and strong Ar II emission band, which observed line may be deleted from a figure. Calculated mode is C-C stretching at all carbon sites.

  Calculated 9.0 $\mu$m band: Accidental overlap with sharp and strong Ar III emission band, which line may be deleted from observed figure. Calculated mode is C-H in-plane bending.

(3) Weaker bands

  **Observed 5.7 $\mu$m band:** Tiny observed band, no calculated peak.

  **Observed 6.0 $\mu$m band:** Tiny observed band, no calculated peak.

  **Observed 11.0 $\mu$m band:** Tiny calculated peak was observed at 10.9 $\mu$m.

  **Observed 12.0 $\mu$m band:** Just same calculated value. Mode is C-H out of plane bending and C-C stretching.

  **Observed 12.7 $\mu$m band:** Calculated was 12.6 $\mu$m. Mode is C-H out of plane bending and C-C stretching.

  **Observed 13.5 $\mu$m band:** Tiny calculated peak at 13.6 $\mu$m. Mode is featured as C-H out of plane bending at pentagons.

  **Observed 14.2 $\mu$m band:** Calculated peak is medium height at 14.0 $\mu$m. Mode is C-H out-of-bending at pentagons.

  **Observed 15.8 $\mu$m band:** Very tiny calculated mark at 15.6 $\mu$m.

  **Observed 16.4 $\mu$m band:** Calculated peak is medium height at 16.5 $\mu$m. Mode is carbon network out-of-plane bending.

  **Observed 17.4 $\mu$m band:** Weak calculated peak at 17.2 $\mu$m.

  **Observed 17.8 $\mu$m band:** Not shown in calculated band, but may be correlated with 18.2 $\mu$m calculated one.

  **Observed 19.0 $\mu$m band:** Calculated peak was 18.8 $\mu$m.

(4) Longer wavelength bands

  Calculated 21.3 $\mu$m band: Need observation data. Mode is hexagon in-plane twisting.

  Calculated 27.6 $\mu$m band: Need observation data.

  Calculated 29.4 $\mu$m band: Need observation data. Mode is hexagon in-plane twisting.

It should be noted that among 16 observed bands we could identify 13 calculated peaks by applying this single molecule.

## 4, CALCULATED INFRARED SPECTRUM OF $(C_{12}H_8)^{3+}$

Calculated infrared spectrum of $(C_{12}H_8)^{3+}$ was illustrated in a bottom of Figure 3 compared with observed NGC2023 on top. Dotted lines are observed IR.

  **Observed 6.2 $\mu$m band**: Calculated value was 6.3 $\mu$m, red shifted by 0.1 $\mu$m, which discrepancy is smaller than $(C_{23}H_{12})^{2+}$. Relative strength is weaker than observed one.

  **Observed 7.7 $\mu$m band:** Calculated band was split into 7.4 and 7.7 $\mu$m, which may be observed as a single strong band.

  **Observed 8.6 $\mu$m band:** Calculated value was the same 8.6 $\mu$m, whereas height is smaller than observed one.

  **Observed 11.2 $\mu$m band:** Calculated value was 11.1 $\mu$m, which relative strength is similar with 7.7 $\mu$m band.

  **Observed 12.7 $\mu$m band:** Calculated value was 12.6 and 12.8 $\mu$m. Strength is smaller than observed one.

  Calculated 3.2 $\mu$m band: Well observed ubiquitous band.

Again, this model molecule became a good example reproducing interstellar infrared spectrum.

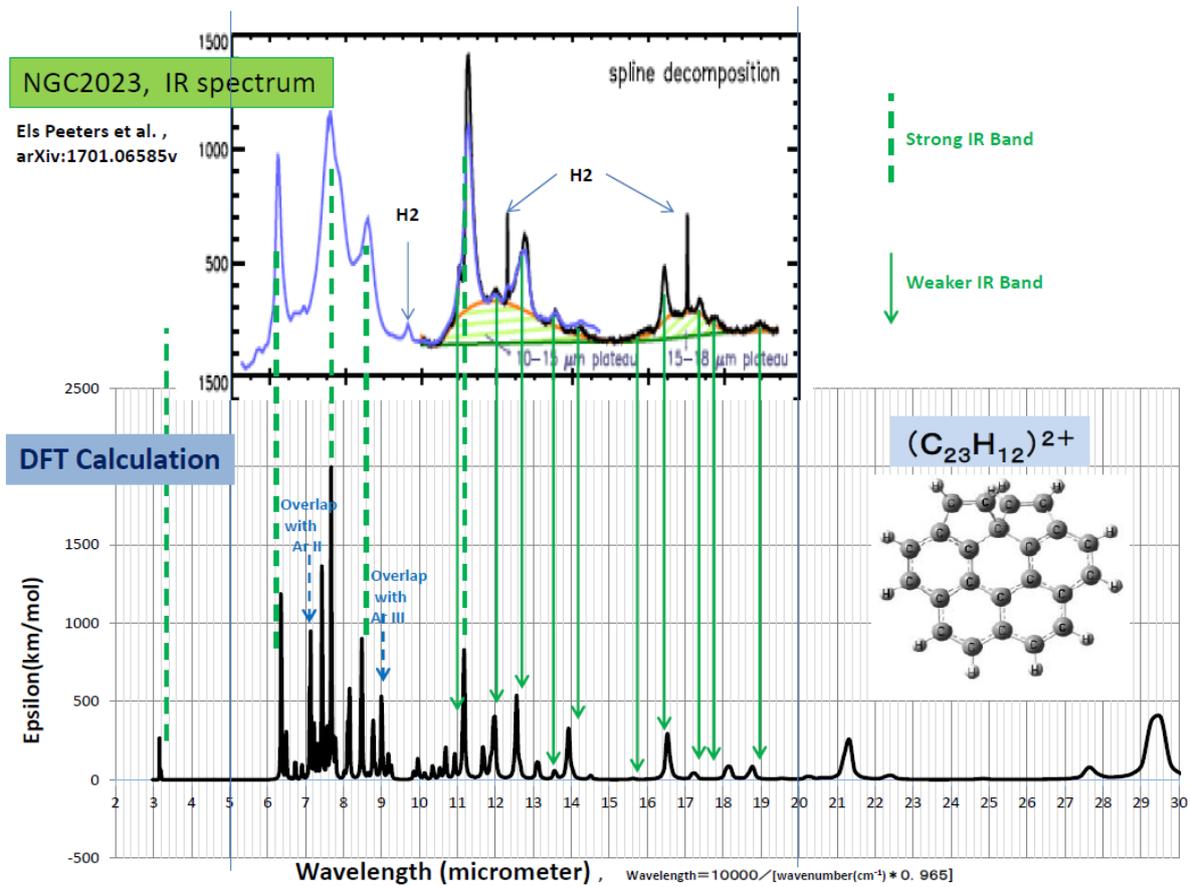

Figure 2, Comparison of infrared spectrum of observed nebula NGC2023 with quantum chemical calculation of hydrocarbon pentagon-hexagon combined molecule $(C_{23}H_{12})^{2+}$. Detailed observed bands were reproduced well by calculation.

Table 1, Observed and calculated infrared spectrum
 Observation by Els Peeters et al
    arXiv.org 1701.06585v1
Calculation by Norio Ota
    This study and
     arXiv.org 1703.05931

| Observed IR of NGC2023 (micrometer) | Calculated IR of $(C23H12)^{2+}$ (micrometer) | Remarks |
|---|---|---|
|  | 3.2 | Well observed band |
| 5.7 |  | Very tiny observed signal |
| 6.0 |  | Very tiny observed signal |
| **6.2** | **6.4** |  |
|  | 7.1 | Overlap with Ar II emission |
|  | 7.5 |  |
| **7.7** | **7.7** |  |
|  | 8.2 |  |
| **8.6** | **8.5** |  |
|  | 8.8 |  |
|  | 9.0 | Overlap with Ar III emission |
| 11.0 | 10.9 |  |
| **11.2** | **11.2** |  |
| 12.0 | 12.0 |  |
| **12.7** | **12.6** |  |
|  | 13.1 |  |
| 13.5 | 13.6 |  |
| 14.2 | 14.0 |  |
| 15.8 | 15.6 | Very tiny calculated peak |
| **16.4** | **16.5** |  |
| 17.4 | 17.2 |  |
| 17.8 |  |  |
|  | 18.2 |  |
| 19.0 | 18.8 |  |
|  | 21.3 | Need observation |
|  | 27.6 | Need observation |
|  | 29.4 | Need observation |

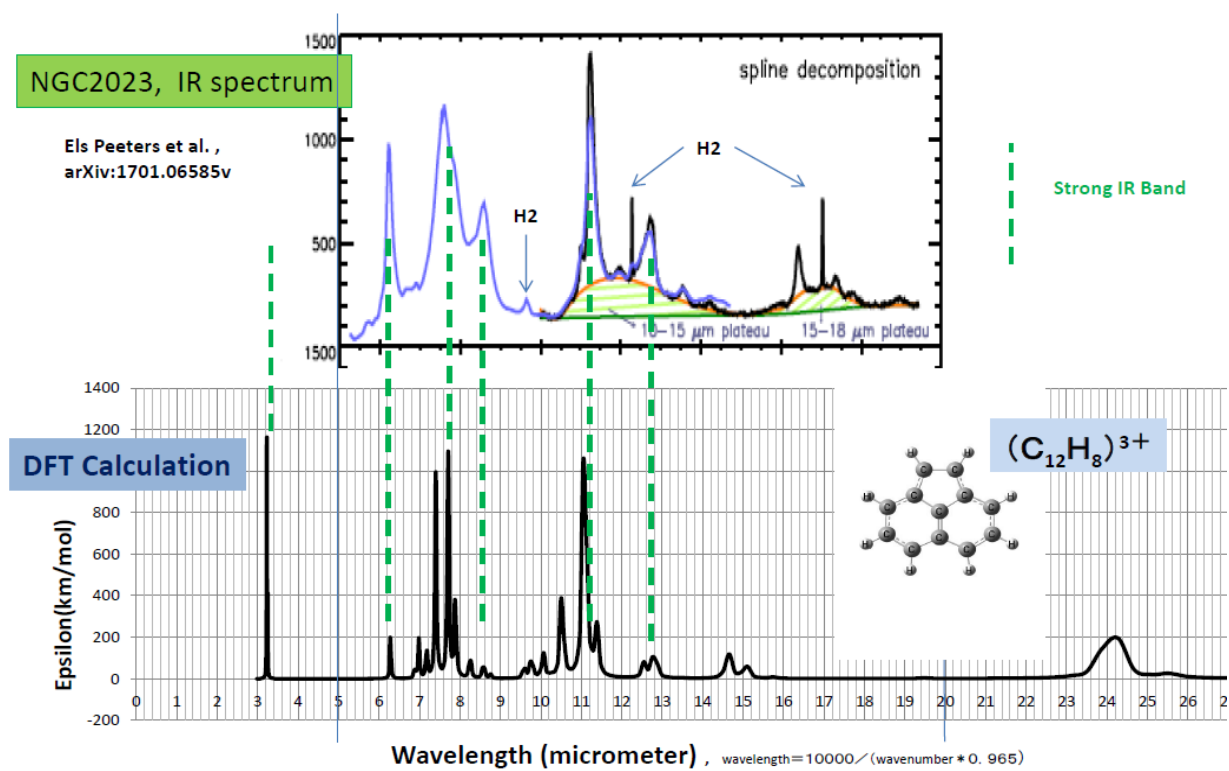

Figure 3, Comparison of infrared spectrum of observed nebula NGC2023 with quantum chemical calculation of tri-cation molecule $(C_{12}H_8)^{3+}$.

6, CONCLUSION

Reflection Nebula NGC2023 shows specific interstellar infrared spectrum due to polycyclic aromatic hydrocarbon (PAH) in a wide wavelength range from 5 to 20 micrometer. In this paper, observed NGC2023 spectrum was reproduced by a quantum chemistry calculation on hydrocarbon pentagon-hexagon combined molecules.
 (1) First example was di-cation $(C_{23}H_{12})^{2+}$ having two hydrocarbon pentagons combined with five hexagons. Observed major bands at 6.2, 7.7, 8.6, and 11.2 micrometer were reproduced well by calculated peaks at 6.4, 7.5, 7.7, 8.5, and 11.2 micrometer.
 (2) Observed weaker bands from 10 to 15 micrometer were 11.0, 12.0, 12.7, 13.5, and 14.2 micrometer, which were computed well as 10.9, 12.0, 12.6, 13.6, and 14.0 micrometer.
 (3) From 15 to 20 micrometer range, observed 15.8, 16.4, 17.4, 17.8, and 19.0 micrometer data were correlated with calculated 15.6, 16.5, 17.2, 18.2, and 18.8 micrometer.
(4) Second model molecule $(C_{12}H_8)^{3+}$ has molecular structure with one hydrocarbon pentagon combined with two hexagons. Observed bands at 6.2, 7.7, 8.6, 11.2 and 12.7 micrometer were calculated well to be 6.3, 7.4, 7.7, 11.1, 12.6, and 12.8 micrometer.
 It was concluded that by introducing hydrocarbon pentagon-hexagon combined molecules, interstellar infrared spectrum could be successfully reproduced in wide wavelength range.

ACKNOWLEDGEMENT

I would like to say great thanks to Els Peeters-sann et al. for providing a very useful data of NGC2023 on open domain arXiv site. These years, internet oriented journal became very important for the progress of worldwide advanced study and quick cooperation.

# REFERENCES

Boersma, C., Bregman, J.D. & Allamandola, L. J.. 2013, ApJ , 769, 117
Boersma, C., Bauschlicher, C. W., Ricca, A., et al. 2014, ApJ Supplement Series, 211:8
Frisch, M. J., Pople, J. A., & Binkley, J. S. 1984, J. Chem. Phys., 80, 3265
Frisch, M. J., Trucks, G.W., Schlegel, H. B., et al. 2009, Gaussian 09, Revision A.02 (Wallingford, CT: Gaussian, Inc.) Geballe,
Nozawa T. et al. 2003, ApJ 598, 785
Nozawa T. et al. , 2006, ApJ 648, 435
Ota, N. 2014a, arXiv org., 1408.6061
Ota, N. 2014b, arXiv org., 1412.0009
Ota, N. 2015a, arXiv org., 1501.01716
Ota, N. 2015b, arXiv org., 1510.07403
Ota, N. 2016, arXiv org., 1603.03399
Ota, N. 2017, arXiv org., 1703.05931
Peeters, Els et al. 2017, arXiv org. 1701.06585v1

Author profile:   Norio Ota, PhD.
 2010~, Senior Professor, University of Tsukuba, Japan
        Spintronics, Optics
1990~2011, Executive officer R&D, Hitachi Maxell Ltd.
        Magnetic storage devices, Optical storage devices
1975~1990, Senior researcher, Hitachi central research laboratory
        Magnetic materials and physics, Optical storage materials
1975, PhD in Physics, Tohoku University, Japan
                Contact e-mail : n-otajitaku@nifty.com
                 Web： https://www.researchgate.net/profile/Norio_Ota_or_Ohta

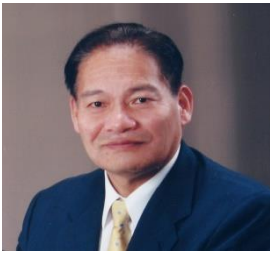

Submitted to arXiv.org on April , 2017 by Norio Ota